\documentclass{article}
\usepackage{spconf,amsmath,graphicx,amsthm,amssymb}
\usepackage{cite} 
\usepackage{multirow}

\title{Noise-robust Speech Recognition with 10 Minutes Unparalleled In-domain Data}
\name{Chen Chen$^1$, Nana Hou$^1$, Yuchen Hu$^1$, Shashank Shirol$^2$, Eng Siong Chng$^1$ 
\sthanks{This research is supported by the Air Traffic Management Research Institute of Nanyang Technological University.}}
\address{$^1$School of Computer Science and Engineering, Nanyang Technological University, Singapore \\  $^2$Manipal Institute of Technology, Manipal, India \\ chen1436@e.ntu.edu.sg}
\begin{document}
%
\maketitle
\begin{abstract}
Noise-robust speech recognition systems require large amounts of training data including noisy speech data and corresponding transcripts to achieve state-of-the-art performances in face of various practical environments. However, such plenty of in-domain data is not always available in the real-life world. In this paper, we propose a generative adversarial network to simulate noisy spectrum from the clean spectrum (Simu-GAN), where only 10 minutes of unparalleled in-domain noisy speech data is required as labels. Furthermore, we also propose a dual-path speech recognition system to improve the robustness of the system under noisy conditions. Experimental results show that the proposed speech recognition system achieves 7.3\% absolute improvement with simulated noisy data by Simu-GAN over the best baseline in terms of word error rate (WER). 
\end{abstract}
\section{Introduction}
\label{sec:intro}

Noise-robust automatic speech recognition (ASR) is a challenging task as huge training data and corresponding transcripts are required to achieve state-of-the-art word error rate (WER) performances \cite{hannun2014deep,chan2016listen,chiu2018state, hu2018generative}. However, such plenty of noisy data is not always available under some practical scenarios as well as collecting target data and transcribing them are also time-consuming and labor-intensive. \par
To address the problem of lacking enough noisy data, prior work \cite{yu2010roles} proposes to train the ASR system with large amounts of clean data and then finetunes with limited noisy in-domain data. Another work \cite{ma2019improving} proposes to extract the pure noise segments from the limited noisy in-domain data and then recursively adds them to the large amounts of clean data to generate the simulated noisy data. Such ``mixup" simulated noisy data are then utilized for the subsequent ASR training.\par
Recent works \cite{goodfellow2014generative,creswell2018generative,isola2017image} in the vision field propose to utilize the generative adversarial network (GAN) to simulate the target-domain images. Furthermore, with only a few minutes of real speech as labels, GAN is also successfully applied in voice conversion task to generate high-quality speech \cite{kameoka2018stargan}. The prior works are the inspiration source of this paper.\par
In this paper, we propose a generative adversarial network to simulate noisy data (Simu-GAN) with only 10 minutes of unparalleled in-domain noisy data for supervision, which provides a promising solution for the noise-robust ASR system with limited in-domain training data. When such clean-to-noisy mapping in the Simu-GAN is trained well, it can generate a large amount of simulated in-domain data at the run-time inference. \par
Specifically, the proposed Simu-GAN consists of a generator and a discriminator. The generator aims to map the clean spectrum to the noisy spectrum and the discriminator is introduced to distinguish the simulated noisy spectrum from the real noisy spectrum \cite{hou2019domain}. At the training stage, the multi-layer patch-wise contrastive loss \cite{park2020contrastive} is utilized to learn the mutual information (\textit{i.e.}, speech content) between the clean spectrum and the simulated noisy spectrum. The discriminator aims to narrow the differences between the simulated noisy spectrum and the real noisy spectrum (\textit{i.e.}, background noises). At the run-time reference, only the generator is required to generate the simulated noisy data.\par

In addition, we propose a dual-path ASR system to evaluate the effectiveness of the proposed Simu-GAN, where both the noisy data and clean data are utilized in the training process. The experiment shows that the proposed Simu-GAN could significantly improve the recognition accuracy and achieve comparable performances compared with the upper-bound baseline.\par

\begin{figure*}[t]
\centering
\includegraphics[width=0.83\textwidth]{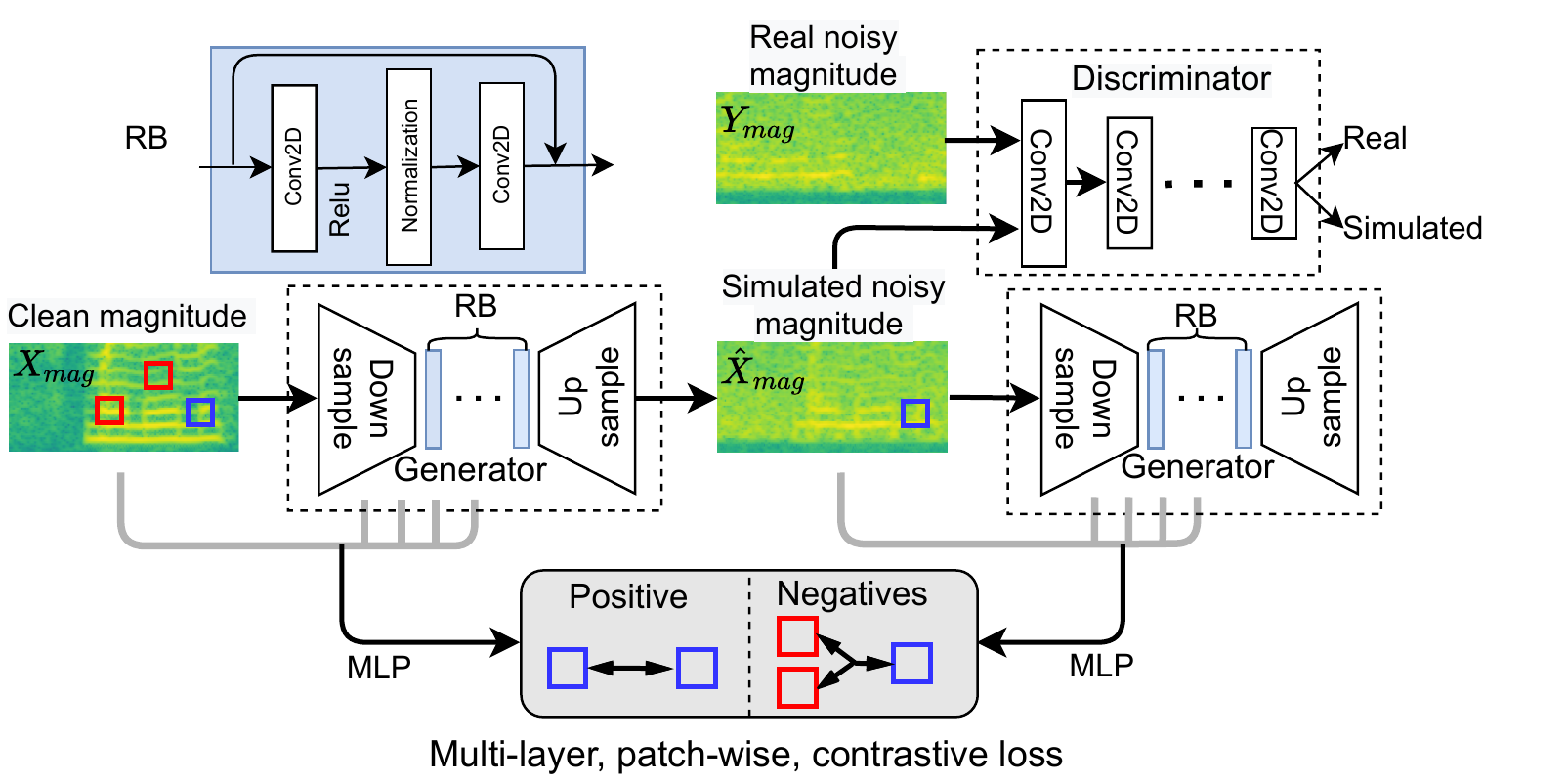}
\vspace{-0.15in}
\caption{The block diagram of proposed Simu-GAN structure. $X$, $\hat{X}$ and $Y$ are the clean features, simulated noisy features and real noisy features, respectively. ``RB" denotes the residual blocks, and ``MLP" denotes two linear layers followed by the ReLU activation function.}    
\vspace{-0.2in}
\label{f1}
\end{figure*}

\section{Simu-GAN ARCHITECTURE}
\label{sec:format}
Given the adequate clean data $X = \{x \in \mathcal{X}\}$ in the source domain and the limited noisy data $Y = \{y \in \mathcal{Y}\}$ in the target domain, we hope to generate the simulated noisy speech $\hat{X}$ that is close to the real noisy speech $Y$. To this end, we propose the Simu-GAN architecture, where the clean speech $X$ and the noisy speech $Y$ are not required to be parallel.
\vspace{-0.1in}
\subsection{Generator and Discriminator}
The generator $G$ is designed to map the clean magnitude $X_{mag}$ to the real noisy magnitude $Y_{mag}$ as shown in Fig.\ref{f1}. We first feed the clean features $X_{mag}$ into two 2-D downsample convolutional layers with the kernel size of (3$\times$3) and the stride of (2$\times$2) as an encoder to learn the embeddings of the input features. Such encoded embeddings are then inputted to nine residual blocks (RB) to learn deep representations. Each residual block includes two convolutional layers with the kernel size of (3$\times$3) and the stride of (1$\times$1) followed by one dropout layer. Finally, two transposed convolutional layers with the kernel size of (3$\times$3) and the stride of (2$\times$2) act as a decoder to upsample the deep representations to the simulated noisy features $\hat{X}_{mag}$. 

We now introduce the discriminator $D$ to distinguish where the inputs come from (\textit{i.e.}, simulated or real). The simulated magnitude features $\hat{X}_{mag}$ and the real noisy magnitude features $Y_{mag}$ are fed into the discriminator, consisting of five 2-D convolutional layers with the kernel size of (4$\times$4) followed by the LeakyReLU activation function. The stride takes (2$\times$2) for the first three convolutional layers and (1$\times$1) for the last two convolutional layers. At the training stage, the adversarial loss \cite{goodfellow2014generative} is employed as:
\begin{small}
\begin{equation}
\mathcal{L}_{GAN} (G,D,X,Y) = \mathbb{E}_{y\sim Y}\log D(y) +\mathbb{E}_{x\sim X}\log (1-D(G(x))).
\end{equation}
\end{small}
By minimizing the adversarial loss, the simulated features $\hat{X}_{mag}$ learn to be visually like the real noisy features $Y_{mag}$. Although the speech content of the simulated and noisy features are different, the discriminator mainly distinguishes the two features by the background noises \cite{hou2021learning}.

\subsection{Multi-layer Patch-wise Contrastive Loss}
As the speech content of the clean and noisy features are different, we introduce the multi-layer patch-wise contrastive loss between the clean features $X_{mag}$ and the simulated features $\hat{X}_{mag}$ to learn the mutual information (\textit{i.e.}, speech content information).

\textbf{Multi-layer and patch-wise}. As shown in Figure \ref{f1}, we reuse the generator $G$ to learn the deep representations of the simulated features $\hat{X}_{mag}$. We first set one small patch in the simulated representations as ``query" and then select the corresponding patch in the clean representations as the positive sample and random 256 patches as the negative samples. Such patches are reshaped via two linear layers with 256 units followed by the ReLU activation function. The contrastive loss is conducted between the positive sample and negative samples from 5 interested layers. 

\textbf{Contrastive Loss}. We now introduce the contrastive learning to learn the mutual information by calculating the cross-entropy loss between the ``query" and the positive/negative patches, formulated as:
\begin{small}
\begin{equation}
\label{lossMPC}
\mathcal{L}_{MPC}(G,X) = \sum_{l=1}^L\sum_{i=1}^{I} -\log\left[ \frac{e^{(\hat{z}_l^i \cdot z_l^i / \tau)}}{e^{(\hat{z}_l^i \cdot z_l^i / \tau)} + \sum_{j=1}^J e^{(\hat{z}_l^i \cdot z_{l}^{j} / \tau)} } \right ]
\end{equation}
\end{small}
where $z_{l}^{i}$ and $\hat{z_{l}^{i}}$ denote the $i^{th}$ positive patches in clean representations and simulated representations of the $l^{th}$ layers in the generator, respectively. $\hat{z_{l}^{j}}$ is the $j^{th}$ negative patches in simulated representations of the $l^{th}$ layers in the generator. $\tau$ presents a temperature parameter in the contrastive learning \cite{wu2018unsupervised}.

Additionally, we also apply the multi-layer patch-wise contrastive loss $\mathcal{L}_{MPC}(G,Y)$ to the real noisy features $Y_{mag}$ to prevent the generator from making unnecessary changes \cite{park2020contrastive}. Therefore, the total loss function $\mathcal{L}_{total}$ of the Simu-GAN is formulated as:
\begin{small}
\begin{equation}
\mathcal{L}_{total} = \mathcal{L}_{GAN}(G,D,X,Y) + \lambda \mathcal{L}_{MPC}(G,X) + \omega \mathcal{L}_{MPC}(G,Y)
\end{equation}
\end{small}
where $\lambda$ and $\omega$ are both set as 1 in this work. 

\begin{figure}[t]
\centering
\includegraphics[width=0.50\textwidth]{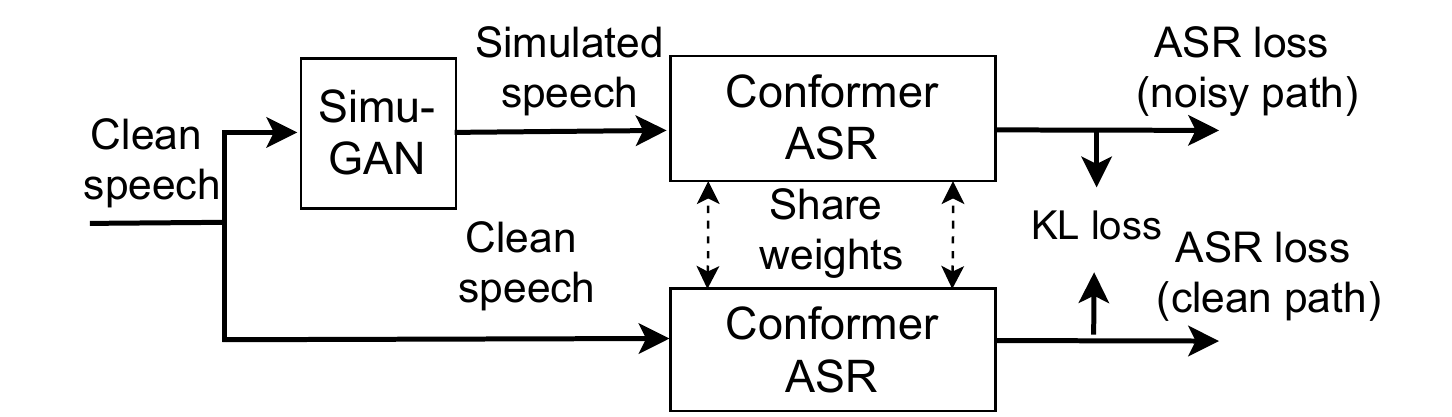}
\vspace{-0.25in}
\caption{The augmented dual-path ASR architecture. Two data flow are fed into Conformer-based ASR network as two independent batches, and the KL loss is computed between the outputs of decoder.}
\vspace{-0.2in}
\label{f3}
\end{figure}

\section{Dual-path ASR System}
To improve the robustness of the ASR systems under noisy conditions, we further propose a dual-path ASR system as shown in Figure \ref{f3}. Specifically, we input the simulated speech generated by the Simu-GAN model and the corresponding clean speech as dual-path inputs into the conformer-based ASR system. We first introduce two ASR losses $\mathcal{L}_{asr}^c$ and $\mathcal{L}_{asr}^n$ calculated for the noisy path and clean path. We then propose a KL divergence-based consistency loss between the two decoder outputs of the noisy path and clean path as $\mathcal{L}_{kl}(X_{dec},\hat{X}_{dec})$. Therefore, the loss of the dual-path ASR system $\mathcal{L}_{dp}$ is formulated as: 
\begin{small}
\begin{equation}
\mathcal{L}_{dp} = \alpha\mathcal{L}_{kl}(X_{dec},\hat{X}_{dec}) + \beta\mathcal{L}_{asr}^c + (1-\beta)\mathcal{L}_{asr}^n 
\end{equation}
\end{small}
where $\alpha, \beta \in [0,1]$ are the parameters to balance the ASR loss $\mathcal{L}_{asr}^{c/n}$ and the KL loss $\mathcal{L}_{kl}(X_{dec},\hat{X}_{dec})$. At the run-time inference, we only utilize the noisy path to evaluate the test set as the clean data of the test set is usually not available.

\section{Experiments and results}
\subsection{Database}
We conduct experiments on the dataset from robust automatic transcription of speech (RATS) program \cite{graff2014rats}, which is recorded with a push-to-talk transceiver by playing back the clean Fisher data. The RATS has eight channels and could provide clean speech, noisy speech, and corresponding transcripts for various training goals. In this work, we select the data in channel A as the in-domain noisy data. The channel A includes 44.3-hours of training data, 4.9-hours of validation data, and 8.2-hours of testing data. 

At the training stage for the proposed Simu-GAN, we only utilize small amounts of unparalleled clean/noisy data from the channel A. At the run-time reference for the Simu-GAN, we use the full clean data of the channel A to generate the simulated noisy data for subsequent ASR training. Such clean data, corresponding simulated noisy data and corresponding transcripts are utilized in the ASR training.

\subsection{Experimental setup}
At the training stage for Simu-GAN, the clean magnitude features were cut into the segments with the dimension of 129 $\times$ 128.  The network was optimized by the Adam algorithm \cite{kingma2014adam} and the learning rate started from 0.002. 

At the training stage for all ASR baselines, the Conformer-based ASR system takes 80-dimensional log-mel feature as the input, where the encoder includes 12 Conformer layers and the decoder consists of 6 Transformer layers \cite{ma2021multitask}. The byte-pair-encoding (BPE) \cite{kudo2018sentencepiece} is utilized as the output token with a size of 994 and the shallow fusion \cite{karita2019comparative} is employed to train a language model with the corresponding transcripts. For the dual-path ASR system, we set the hyper-parameters $\alpha$ and $\beta$ to 0.4 and 0.7, respectively. 

\subsection{Reference Baselines}
To evaluate the effectiveness of the proposed Simu-GAN and the dual-path ASR system, we built 5 baselines for comparison.
\begin{itemize}
    \item Clean-ASR \cite{ma2021multitask}: we train the single-path Conformer-based ASR system only with the clean RATS data. 
    \item SpecAugment \cite{park2019specaugment}: we swap the one frequency bin and one frame as a data augmentation approach for the clean-ASR system training. 
    \item Finetune \cite{yu2010roles}: we tune the pre-trained clean-ASR system with the 10 minutes RATS channel A data. 
    \item Mixup \cite{ma2019improving}: we generate the simulated noisy data by adding the noises segments extracting from full RATS channel A data and then train the single-path Conformer-based ASR system with the ``mixup" simulated data. 
    \item Noisy-ASR \cite{ma2021multitask}: we train the single-path Conformer-based ASR system with the real RATS channel A data, which is the upper-bound performances with full in-domain data. 
\end{itemize}

\subsection{Results}

\subsubsection{Effect of the different amount of data for Simu-GAN training}
We first analyze and summarize the WER performances of Simu-GAN with different amounts of training data on the single-path ASR system. Specifically, we utilize the different amounts of clean/noisy unparalleled data for the Simu-GAN training, and then the approximate 50-hours clean data (44.3+4.9 hours) of RATS channel A are utilized to generate the simulated in-domain data for the subsequent ASR training and evaluation. 

\begin{figure}[t]
\centering
\includegraphics[width=0.48\textwidth]{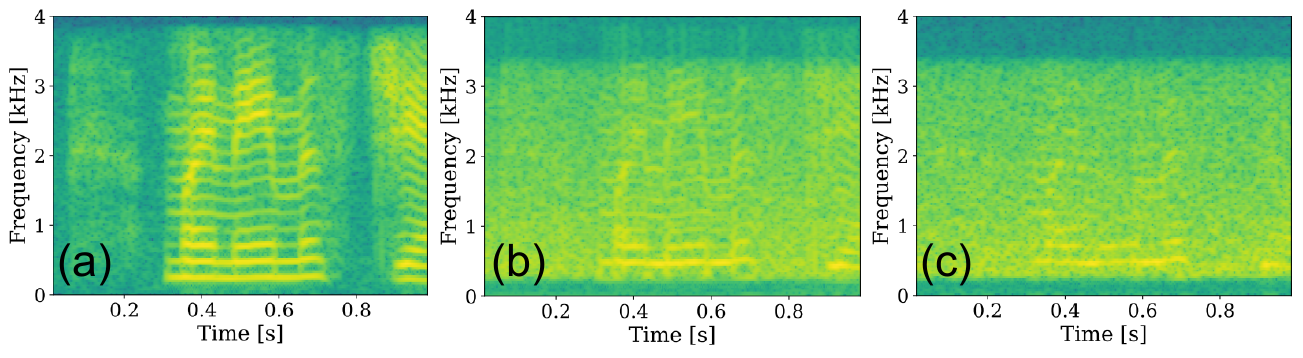}
\vspace{-0.28in}
\caption{The magnitude of a sample (10315\_21016.wav) for (a) clean magnitude, (b) simulated noisy magnitude by Simu-GAN, and (c) real noisy magnitude from RATS channel A (ground-truth).}
\vspace{-0.28in}
\label{f4}
\end{figure}

From Table \ref{table3}, we observe that the performances significantly improve as the amount of clean/noisy training data increases. We obtain the best WER of 68.9\% for the single-path ASR system with 1-hour clean training data and 10 minutes of noisy training data. To further show the contribution of the proposed Simu-GAN approach (system 5), we illustrate the magnitude spectrum of an example as shown in Figure \ref{f4}. We can see that the proposed Simu-GAN approach can produce approximately the same spectrum (seen in Figure \ref{f4}(b)) with the real RATS channel A sample (seen in Figure \ref{f4}(c)). More listening samples are available at Github\footnote{https://chrisole.github.io/ICASSP2022-demo/}.

\begin{table}[t]
\centering
\caption{The comparative study of the different amount of data for the Simu-GAN training. The WER (\%) performances are evaluated on the single-path ASR system using the real test set of RATS channel A. ``Clean speech" denotes the amount of the clean data of RATS channel A and ``Noisy speech" presents the amount of the RATS channel A data. No speed perturbation is utilized in the following experiments.}
\vspace{0.08in}
\resizebox{0.38\textwidth}{!}{
\begin{tabular}{c|c|c|c}
\hline\hline
{\begin{tabular}[c]{@{}c@{}}System\\ No.\end{tabular}} &
  \multicolumn{2}{c|}{Data for Simu-GAN training} &
 {\begin{tabular}[c]{@{}c@{}}WER\\ (\%)\end{tabular}} \\ \cline{2-3}
 &
  Clean speech &
  Noisy speech &
   \\ \hline\hline
1 & 1 min  &  1 min  & 87.2 \\ \hline
2 & 2 min  &  2 min  & 74.2 \\ \hline
3 & 5 min  &  5 min  & 72.7 \\ \hline
4 & 10 min &  10 min  & 72.4 \\ \hline
5 & 1 h  &    10 min    & \textbf{68.9} \\ \hline
  \hline
\end{tabular}}
\vspace{-0.1in}
\label{table3}
\end{table}

\vspace{-0.1in}
\subsubsection{Effect of the proposed dual-path ASR system}
We further report the effect of the proposed dual-path ASR system trained by the simulated noisy data, corresponding clean data, and transcripts shown in Table \ref{table4}. Such simulated noisy data are generated by the proposed Simu-GAN with different amounts of training data. We observe that the proposed dual-path ASR systems could further improve the performances compared with the single-path ASR systems under the same amounts of training data. We obtain the best WER of 60.7\% on the proposed dual-path ASR system with simulated data by Simu-GAN, which is trained by 1-hour clean training data and 10 minutes unparalleled noisy training data. We adopt this setting for Simu-GAN and the dual-path ASR system hereafter.

\begin{table}[t]
\centering
\caption{The comparative study of the proposed dual-path ASR systems. ``\# ASR path" presents the number of path of the ASR systems (single-path or dual-path). ``S.P." denotes the speed perturbation with $\{\times 0.9, \times 1.0, \times 1.1\}$.}
\vspace{0.08in}
\resizebox{0.48\textwidth}{!}{
\begin{tabular}{c|c|c|c|c|c}
\hline\hline
{\begin{tabular}[c]{@{}c@{}}System\\ No.\end{tabular}} &
 \multicolumn{2}{|c|}{Data for Simu-GAN training} &
{\begin{tabular}[c]{@{}c@{}} \# ASR path\end{tabular}} &
{S.P.} &
{\begin{tabular}[c]{@{}c@{}}WER\\ (\%)\end{tabular}} \\ \cline{2-3}
  & Clean data & Noisy data &        &     &      \\ \hline\hline
\multirow{2}{*}{4} & 10 min     & 10 min     & Single & $\times$  & 72.4 \\ \cline{2-6}
                   & 10 min     & 10 min     & Single & $\surd$ & 67.0   \\ \hline
\multirow{2}{*}{5} & 1 h        & 10 min     & Single   & $\times$ &  68.9 \\ \cline{2-6}                 
                   & 1 h        & 10 min     & Single   & $\surd$ &  63.0 \\
\hline\hline
\multirow{2}{*}{6} & 10 min     & 10 min     & Dual    & $\times$  & 65.4 \\ \cline{2-6}
                   & 10 min     & 10 min     & Dual    & $\surd$ & 61.5 \\ \hline
\multirow{2}{*}{7} & 1 h        & 10 min     & Dual   & $\times$ &  65.4 \\ \cline{2-6}
                   & 1 h        & 10 min     & Dual   & $\surd$ &  \textbf{60.7} \\ \hline\hline
\end{tabular}}
\vspace{-0.15in}
\label{table4}
\end{table}

\vspace{-0.1in}
\subsubsection{Benchmark against other competitive methods}
Table \ref{table5} summarizes the comparison between the proposed Simu-GAN and other competitive techniques in terms of the WER (\%). We observe that the proposed Simu-GAN obtained the best performance. Comparing with the ``Mixup" methods, the proposed Simu-GAN achieves the 7.3\% absolute WER improvements. In addition, our best performance is also close to the upper-bound baseline trained by real in-domain data.

\begin{table}[]
\centering
\caption{The comparative study of other competitive techniques. Speed perturbation with $\{\times 0.9, \times 1.0, \times 1.1\}$ are utilized in the following experiments.}
\vspace{0.08in}
\resizebox{0.49\textwidth}{!}{
\begin{tabular}{c|c|c|c|c}
\hline\hline
{\begin{tabular}[c]{@{}c@{}}Method\\ Name\end{tabular}} &
  \multicolumn{3}{c|}{All data requirements for training} &
{\begin{tabular}[c]{@{}c@{}}WER\\ (\%)\end{tabular}} \\ \cline{2-4}
 &
  Clean speech &
  Real noisy speech &
  Simulated noisy speech
   &
   \\ \hline\hline
Clean-ASR &
  44.3 h &
  - & - &
  93.4 \\ \hline
  \begin{tabular}[c]{@{}c@{}}SpecAugment \cite{park2019specaugment}\\ (slight)\end{tabular} &
  44.3 h & - &
  - &

  84.7 \\ \hline
Finetune \cite{yu2010roles} &
  44.3 h &
  \begin{tabular}[c]{@{}c@{}}10 min \\ with label\end{tabular} & - &
  73.0 \\ \hline
  Mixup \cite{ma2019improving} &
  44.3 h & 
  \begin{tabular}[c]{@{}c@{}}44.3 h \\ without label\end{tabular}& 44.3 h &
  68.0 \\ \hline
  Simu-GAN (ours)&
  44.3 h &
  \begin{tabular}[c]{@{}c@{}}10 min \\ without label\end{tabular} & 44.3 h &
  \textbf{60.7} \\ \hline\hline
  Noisy-ASR &
  - &
  \begin{tabular}[c]{@{}c@{}}44.3 h\\ with label\end{tabular} & - &
  49.5 \\ \hline\hline
\end{tabular}}
\vspace{-0.2in}
\label{table5}
\end{table}

\section{Conclusion}
We propose a generative adversarial network to simulate the noisy in-domain speech from the clean speech (Simu-GAN) with 10 minutes of real noisy samples as labels to address the problem that the in-domain data is limited. We also propose a dual-path ASR system to improve the robustness of the ASR systems under noisy conditions. Experimental results show that the proposed Simu-GAN achieves the 7.3\% absolute WER improvements on the dual-path ASR system compared with the best baseline.

\vfill\pagebreak


\bibliographystyle{IEEEtran}
\bibliography{refs}

\begin{thebibliography}{10}
\providecommand{\url}[1]{#1}
\csname url@samestyle\endcsname
\providecommand{\newblock}{\relax}
\providecommand{\bibinfo}[2]{#2}
\providecommand{\BIBentrySTDinterwordspacing}{\spaceskip=0pt\relax}
\providecommand{\BIBentryALTinterwordstretchfactor}{4}
\providecommand{\BIBentryALTinterwordspacing}{\spaceskip=\fontdimen2\font plus
\BIBentryALTinterwordstretchfactor\fontdimen3\font minus
  \fontdimen4\font\relax}
\providecommand{\BIBforeignlanguage}[2]{{%
\expandafter\ifx\csname l@#1\endcsname\relax
\typeout{** WARNING: IEEEtran.bst: No hyphenation pattern has been}%
\typeout{** loaded for the language `#1'. Using the pattern for}%
\typeout{** the default language instead.}%
\else
\language=\csname l@#1\endcsname
\fi
#2}}
\providecommand{\BIBdecl}{\relax}
\BIBdecl

\bibitem{hannun2014deep}
A.~Hannun, C.~Case, J.~Casper, B.~Catanzaro, G.~Diamos, E.~Elsen, R.~Prenger,
  S.~Satheesh, S.~Sengupta, A.~Coates \emph{et~al.}, ``Deep speech: Scaling up
  end-to-end speech recognition,'' \emph{arXiv preprint arXiv:1412.5567}, 2014.

\bibitem{chan2016listen}
W.~Chan, N.~Jaitly, Q.~Le, and O.~Vinyals, ``Listen, attend and spell: A neural
  network for large vocabulary conversational speech recognition,'' in
  \emph{2016 IEEE International Conference on Acoustics, Speech and Signal
  Processing (ICASSP)}.\hskip 1em plus 0.5em minus 0.4em\relax IEEE, 2016, pp.
  4960--4964.

\bibitem{chiu2018state}
C.-C. Chiu, T.~N. Sainath, Y.~Wu, R.~Prabhavalkar, P.~Nguyen, Z.~Chen,
  A.~Kannan, R.~J. Weiss, K.~Rao, E.~Gonina \emph{et~al.}, ``State-of-the-art
  speech recognition with sequence-to-sequence models,'' in \emph{2018 IEEE
  International Conference on Acoustics, Speech and Signal Processing
  (ICASSP)}.\hskip 1em plus 0.5em minus 0.4em\relax IEEE, 2018, pp. 4774--4778.

\bibitem{hu2018generative}
H.~Hu, T.~Tan, and Y.~Qian, ``Generative adversarial networks based data
  augmentation for noise robust speech recognition,'' in \emph{2018 IEEE
  International Conference on Acoustics, Speech and Signal Processing
  (ICASSP)}.\hskip 1em plus 0.5em minus 0.4em\relax IEEE, 2018, pp. 5044--5048.

\bibitem{yu2010roles}
D.~Yu, L.~Deng, and G.~Dahl, ``Roles of pre-training and fine-tuning in
  context-dependent dbn-hmms for real-world speech recognition,'' in
  \emph{Proc. NIPS Workshop on Deep Learning and Unsupervised Feature
  Learning}, 2010.

\bibitem{ma2019improving}
D.~Ma, G.~Li, H.~Xu, and E.~S. Chng, ``Improving code-switching speech
  recognition with data augmentation and system combination,'' in \emph{2019
  Asia-Pacific Signal and Information Processing Association Annual Summit and
  Conference (APSIPA ASC)}.\hskip 1em plus 0.5em minus 0.4em\relax IEEE, 2019,
  pp. 1308--1312.

\bibitem{goodfellow2014generative}
I.~Goodfellow, J.~Pouget-Abadie, M.~Mirza, B.~Xu, D.~Warde-Farley, S.~Ozair,
  A.~Courville, and Y.~Bengio, ``Generative adversarial nets,'' \emph{Advances
  in neural information processing systems}, vol.~27, 2014.

\bibitem{creswell2018generative}
A.~Creswell, T.~White, V.~Dumoulin, K.~Arulkumaran, B.~Sengupta, and A.~A.
  Bharath, ``Generative adversarial networks: An overview,'' \emph{IEEE Signal
  Processing Magazine}, vol.~35, no.~1, pp. 53--65, 2018.

\bibitem{isola2017image}
P.~Isola, J.-Y. Zhu, T.~Zhou, and A.~A. Efros, ``Image-to-image translation
  with conditional adversarial networks,'' in \emph{Proceedings of the IEEE
  conference on computer vision and pattern recognition}, 2017, pp. 1125--1134.

\bibitem{kameoka2018stargan}
H.~Kameoka, T.~Kaneko, K.~Tanaka, and N.~Hojo, ``Stargan-vc: Non-parallel
  many-to-many voice conversion using star generative adversarial networks,''
  in \emph{2018 IEEE Spoken Language Technology Workshop (SLT)}.\hskip 1em plus
  0.5em minus 0.4em\relax IEEE, 2018, pp. 266--273.

\bibitem{hou2019domain}
N.~Hou, C.~Xu, E.~S. Chng, and H.~Li, ``Domain adversarial training for speech
  enhancement,'' in \emph{2019 Asia-Pacific Signal and Information Processing
  Association Annual Summit and Conference (APSIPA ASC)}.\hskip 1em plus 0.5em
  minus 0.4em\relax IEEE, 2019, pp. 667--672.

\bibitem{park2020contrastive}
T.~Park, A.~A. Efros, R.~Zhang, and J.-Y. Zhu, ``Contrastive learning for
  unpaired image-to-image translation,'' in \emph{European Conference on
  Computer Vision}.\hskip 1em plus 0.5em minus 0.4em\relax Springer, 2020, pp.
  319--345.

\bibitem{hou2021learning}
N.~Hou, C.~Xu, E.~S. Chng, and H.~Li, ``Learning disentangled feature
  representations for speech enhancement via adversarial training,'' in
  \emph{ICASSP 2021-2021 IEEE International Conference on Acoustics, Speech and
  Signal Processing (ICASSP)}.\hskip 1em plus 0.5em minus 0.4em\relax IEEE,
  2021, pp. 666--670.

\bibitem{wu2018unsupervised}
Z.~Wu, Y.~Xiong, S.~X. Yu, and D.~Lin, ``Unsupervised feature learning via
  non-parametric instance discrimination,'' in \emph{Proceedings of the IEEE
  conference on computer vision and pattern recognition}, 2018, pp. 3733--3742.

\bibitem{graff2014rats}
D.~Graff, K.~Walker, S.~M. Strassel, X.~Ma, K.~Jones, and A.~Sawyer, ``The rats
  collection: Supporting hlt research with degraded audio data.'' in
  \emph{LREC}.\hskip 1em plus 0.5em minus 0.4em\relax Citeseer, 2014, pp.
  1970--1977.

\bibitem{kingma2014adam}
D.~P. Kingma and J.~Ba, ``Adam: A method for stochastic optimization,''
  \emph{arXiv preprint arXiv:1412.6980}, 2014.

\bibitem{ma2021multitask}
D.~Ma, N.~Hou, V.~T. Pham, H.~Xu, and E.~S. Chng, ``Multitask-based joint
  learning approach to robust asr for radio communication speech,'' \emph{arXiv
  preprint arXiv:2107.10701}, 2021.

\bibitem{kudo2018sentencepiece}
T.~Kudo and J.~Richardson, ``Sentencepiece: A simple and language independent
  subword tokenizer and detokenizer for neural text processing,'' \emph{arXiv
  preprint arXiv:1808.06226}, 2018.

\bibitem{karita2019comparative}
S.~Karita, N.~Chen, T.~Hayashi, T.~Hori, H.~Inaguma, Z.~Jiang, M.~Someki,
  N.~E.~Y. Soplin, R.~Yamamoto, X.~Wang \emph{et~al.}, ``A comparative study on
  transformer vs rnn in speech applications,'' in \emph{2019 IEEE Automatic
  Speech Recognition and Understanding Workshop (ASRU)}.\hskip 1em plus 0.5em
  minus 0.4em\relax IEEE, 2019, pp. 449--456.

\bibitem{park2019specaugment}
D.~S. Park, W.~Chan, Y.~Zhang, C.-C. Chiu, B.~Zoph, E.~D. Cubuk, and Q.~V. Le,
  ``Specaugment: A simple data augmentation method for automatic speech
  recognition,'' \emph{arXiv preprint arXiv:1904.08779}, 2019.

\end{thebibliography}

\end{document}